\documentclass[twocolumn,showpacs,preprintnumbers,amsmath,aps]{revtex4}
\usepackage{graphicx}
\usepackage{dcolumn}
\usepackage{bm}
\usepackage{color}
\begin{document}
%
\def\eq#1{(\ref{#1})}
\def\fig#1{Fig.\hspace{1mm}\ref{#1}}
\def\tab#1{\hspace{1mm}\ref{#1}}
%
\title{
The superconducting state in the $\rm{B_{2}H_{6}}$ compound at $360$ GPa}
\author{R. Szcz{\c{e}}{\'s}niak, E.A. Drzazga, A.M. Duda}
\affiliation{Institute of Physics, Cz{\c{e}}stochowa University of Technology, Al. Armii Krajowej 19, 42-200 Cz{\c{e}}stochowa, Poland}
\email{aduda@wip.pcz.pl}
\date{\today} 
\begin{abstract}
In the paper, the thermodynamic properties of the superconducting state in the $\rm{B_{2}H_{6}}$ compound have been characterized. The pressure of $360$ GPa has been taken into account. The calculations have been carried out in the framework of the Eliashberg formalism for the wide range of the Coulomb pseudopotential: $\mu^{\star}\in\left<0.1,0.3\right>$. It has been found that the critical temperature ($T_{C}$) varies in the range from $147$ K to 
$87$ K, depending on the assumed value of the Coulomb pseudopotential. The ratio of the energy gap to the critical temperature 
($R_{\Delta}\equiv 2\Delta\left(0\right)/k_{B}T_{C}$) significantly exceeds the value predicted by the BCS theory: $R_{\Delta}\in\left<4.24,3.98\right>$. In the similar manner behaves the ratio of the specific heat jump to the heat of the normal state ($R_{C}\equiv\Delta C\left(T_{C}\right)/C^{N}\left(T_{C}\right)$), namely: $R_{C}\in\left<2.33,2.17\right>$. The parameter 
$R_{H}\equiv T_{C}C^{N}\left(T_{C}\right)/H^{2}_{C}\left(0\right)$, where $H_{C}\left(0\right)$ is the thermodynamic critical field, ranges from $0.144$ to $0.168$. 
\end{abstract}
\maketitle

\vspace*{1cm}


The intensive researches on the properties of the superconducting state have lasted for over a hundred years \cite{Onnes}. Their main goal is to get the material in which the superconducting state would exist at the room temperature. So far, the highest critical temperature ($T_{C}$) equal to $164$ K has been measured in the ${\rm HgBa_{2}Ca_{2}Cu_{3}O_{8+y}}$ superconductor, located under the pressure at $\sim 31$ GPa \cite{GaoHTSC}. However, due to the lack of the acceptable theory of the superconducting state in cuprates, it is difficult to answer the question whether in the present group of the materials might be possible to get even higher value of $T_{C}$ \cite{Radek01}. 

In 1968, Ashcroft has suggested that the metallic hydrogen under the action of the high pressure would be the room-temperature superconductor \cite{Ashcroft}. Later, after the complicated numerical calculations, it has been found that the pressure of the metallization ($p_{m}$) for the hydrogen molecular phase equals about $400$ GPa \cite{Stadele}. Thus, in the questioned pressure range, the existence of the superconducting state with the very high critical temperature can be expected \cite{Cudazzo}, \cite{Zhang}, \cite{Radek02}, \cite{Radek03}. 

Increasing in the value of the pressure up to $\sim 500$ GPa causes the dissociation of the hydrogen into the atomic phase. It should be noted that near the pressure of the molecular dissociation ($p=539$ GPa), the critical temperature can be equal to $360$ K \cite{Radek04}.

Currently, the highest value of $T_{C}$ for the metallic hydrogen is expected for the pressure at $2$ TPa. In the considered case, the critical temperature can reach up $600$ K \cite{Maksimov}, \cite{Radek05}, \cite{McMahon}.    

Due to the fact that the superconducting state in the metallic hydrogen can be formed only for very high pressures, scientists began to look for other physical system, in which $T_{C}$ takes high value, while $p_{m}$ is relatively low. Based on the survey, it has been found that the most interesting group is the family of the hydrogen-rich compounds \cite{Ashcroft1}, \cite{Tse}, \cite{Gao}, \cite{Canales}, \cite{Gao1}, \cite{Chen}, \cite{Eremets}. For example, in the silicon compounds ${\rm Si_{2}H_{6}}$ ($p = 275$ GPa) and ${\rm SiH_{4}\left(H_{2}\right)_{2}}$ ($p = 250$ GPa), the maximum of the critical temperature takes the values of $173$ K and $130$ K, respectively \cite{Radek06}, \cite{Radek07}. 


In the present paper, we have described the results obtained for the superconducting state in the $\rm{B_{2}H_{6}}$ compound ($p=360$ GPa) \cite{Kazutaka}. Due to the high value of the electron-phonon coupling constant ($\lambda=1.32$), the calculations have been carried out in the framework of the Eliashberg formalism \cite{Eliashberg}. It should be noted that this formalism is the natural generalization of the classical BCS theory \cite{BCS}. 

On the imaginary axis ($i\equiv\sqrt{-1}$), the Eliashberg equations create the infinite system of the non-linear algebraic equations with the integral kernel, which allows to determine the order parameter function ($\phi_{n}\equiv\phi\left(i\omega_{n}\right)$) and the wave function renormalization factor 
($Z_{n}\equiv Z\left(i\omega_{n}\right)$). The quantity $\omega_{n}$ denotes the n-th Matsubara frequency: 
$\omega_{n}\equiv \left(\pi / \beta\right)\left(2n-1\right)$, where $\beta\equiv\left(k_{B}T\right)^{-1}$ ($k_{B}$ is the Boltzmann constant). The order parameter is defined by the ratio: $\Delta_{n}\equiv \phi_{n}/Z_{n}$. 

The open form of the Eliashberg equations can be written as:

\begin{equation}
\label{r1}
\phi_{n}=\frac{\pi}{\beta}\sum_{m=-M}^{M}
\frac{\lambda\left(i\omega_{n}-i\omega_{m}\right)-\mu^{\star}\theta\left(\omega_{c}-|\omega_{m}|\right)}
{\sqrt{\omega_m^2Z^{2}_{m}+\phi^{2}_{m}}}\phi_{m},
\end{equation}
\begin{equation}
\label{r2}
Z_{n}=1+\frac{1}{\omega_{n}}\frac{\pi}{\beta}\sum_{m=-M}^{M}
\frac{\lambda\left(i\omega_{n}-i\omega_{m}\right)}{\sqrt{\omega_m^2Z^{2}_{m}+\phi^{2}_{m}}}\omega_{m}Z_{m},
\end{equation}
where the pairing kernel for the electron-phonon interaction is given by the formula:
\begin{equation}
\label{r3}
\lambda\left(z\right)\equiv 2\int_0^{\Omega_{\rm{max}}}d\Omega\frac{\Omega}{\Omega ^2-z^{2}}\alpha^{2}F\left(\Omega\right).
\end{equation}
The spectral function ($\alpha^{2}F\left(\Omega\right)$) for the $\rm{B_{2}H_{6}}$ compound under the pressure at $360$ GPa has been calculated in the paper \cite{Kazutaka}. The maximum phonon frequency is equal to $368.5$ meV.

The parameter $\mu^{\star}$ is called the Coulomb pseudopotential and serves for the modeling of the depairing Coulomb correlations \cite{Morel}; $\theta$ denotes the Heaviside unit function, and $\omega_{c}$ is the cut-off frequency; $\omega_{c}=3\Omega_{\rm{max}}$.

When carrying out the numerical calculations the finite number of the equations should be taken into account. It turns out that above simplification does not make the significant error to the final result, if the value of the considered temperature is not too low. In the study, we have adopted $M=1100$ and $T\in\left<T_{0}=25{\rm K}, T_{C}\right>$, which ensured the convergence of the functions $\phi_{n}$ and $Z_{n}$.

It is worth noting that the Eliashberg equations have been solved using the numerical procedures, which have been tested and discussed in the papers \cite{Radek08} and \cite{Radek09}.

From the physical point of view, the Eliashberg equations on the imaginary axis allow to determine the critical temperature, the free energy, the thermodynamic critical field, and the specific heat of the superconducting state. However, with their help it is impossible to accurately calculate the physical value of the energy gap ($2\Delta$) and the electron effective mass ($m^{\star}_{e}$). To estimate the parameters $2\Delta$ and $m^{\star}_{e}$, the solutions of the Eliashberg equations should be analytically continued from the imaginary axis to the real axis ($\omega$). Such is the purpose of the Eliashberg equations in the mixed representation \cite{Marsiglio}:

%
\begin{widetext}
\begin{eqnarray}
\label{r4}
\phi\left(\omega+i\delta\right)&=&
                                  \frac{\pi}{\beta}\sum_{m=-M}^{M}
                                  \left[\lambda\left(\omega-i\omega_{m}\right)-\mu^{\star}\theta\left(\omega_{c}-|\omega_{m}|\right)\right]
                                  \frac{\phi_{m}}
                                  {\sqrt{\omega_m^2Z^{2}_{m}+\phi^{2}_{m}}}\\ \nonumber
                              &+& i\pi\int_{0}^{+\infty}d\omega^{'}\alpha^{2}F\left(\omega^{'}\right)
                                  \left[\left[N\left(\omega^{'}\right)+f\left(\omega^{'}-\omega\right)\right]
                                  \frac{\phi\left(\omega-\omega^{'}+i\delta\right)}
                                  {\sqrt{\left(\omega-\omega^{'}\right)^{2}Z^{2}\left(\omega-\omega^{'}+i\delta\right)
                                  -\phi^{2}\left(\omega-\omega^{'}+i\delta\right)}}\right]\\ \nonumber
                              &+& i\pi\int_{0}^{+\infty}d\omega^{'}\alpha^{2}F\left(\omega^{'}\right)
                                  \left[\left[N\left(\omega^{'}\right)+f\left(\omega^{'}+\omega\right)\right]
                                  \frac{\phi\left(\omega+\omega^{'}+i\delta\right)}
                                  {\sqrt{\left(\omega+\omega^{'}\right)^{2}Z^{2}\left(\omega+\omega^{'}+i\delta\right)
                                  -\phi^{2}\left(\omega+\omega^{'}+i\delta\right)}}\right],
\end{eqnarray}
and
\begin{eqnarray}
\label{r5}
Z\left(\omega+i\delta\right)&=&
                                  1+\frac{i}{\omega}\frac{\pi}{\beta}\sum_{m=-M}^{M}
                                  \lambda\left(\omega-i\omega_{m}\right)
                                  \frac{\omega_{m}Z_{m}}
                                  {\sqrt{\omega_m^2Z^{2}_{m}+\phi^{2}_{m}}}\\ \nonumber
                              &+&\frac{i\pi}{\omega}\int_{0}^{+\infty}d\omega^{'}\alpha^{2}F\left(\omega^{'}\right)
                                  \left[\left[N\left(\omega^{'}\right)+f\left(\omega^{'}-\omega\right)\right]
                                  \frac{\left(\omega-\omega^{'}\right)Z\left(\omega-\omega^{'}+i\delta\right)}
                                  {\sqrt{\left(\omega-\omega^{'}\right)^{2}Z^{2}\left(\omega-\omega^{'}+i\delta\right)
                                  -\phi^{2}\left(\omega-\omega^{'}+i\delta\right)}}\right]\\ \nonumber
                              &+&\frac{i\pi}{\omega}\int_{0}^{+\infty}d\omega^{'}\alpha^{2}F\left(\omega^{'}\right)
                                  \left[\left[N\left(\omega^{'}\right)+f\left(\omega^{'}+\omega\right)\right]
                                  \frac{\left(\omega+\omega^{'}\right)Z\left(\omega+\omega^{'}+i\delta\right)}
                                  {\sqrt{\left(\omega+\omega^{'}\right)^{2}Z^{2}\left(\omega+\omega^{'}+i\delta\right)
                                  -\phi^{2}\left(\omega+\omega^{'}+i\delta\right)}}\right], 
\end{eqnarray}
\end{widetext}
%
where $N\left(\omega\right)$ and $f\left(\omega\right)$ denote the Bose-Einstein and Fermi-Dirac functions, respectively. 

Note that the Eliashberg equations in the mixed representation have been solved using the procedures discussed and tested in the papers \cite{Radek10} and \cite{Radek11}.


%
\begin{figure*}[ht]
\includegraphics[scale=0.65]{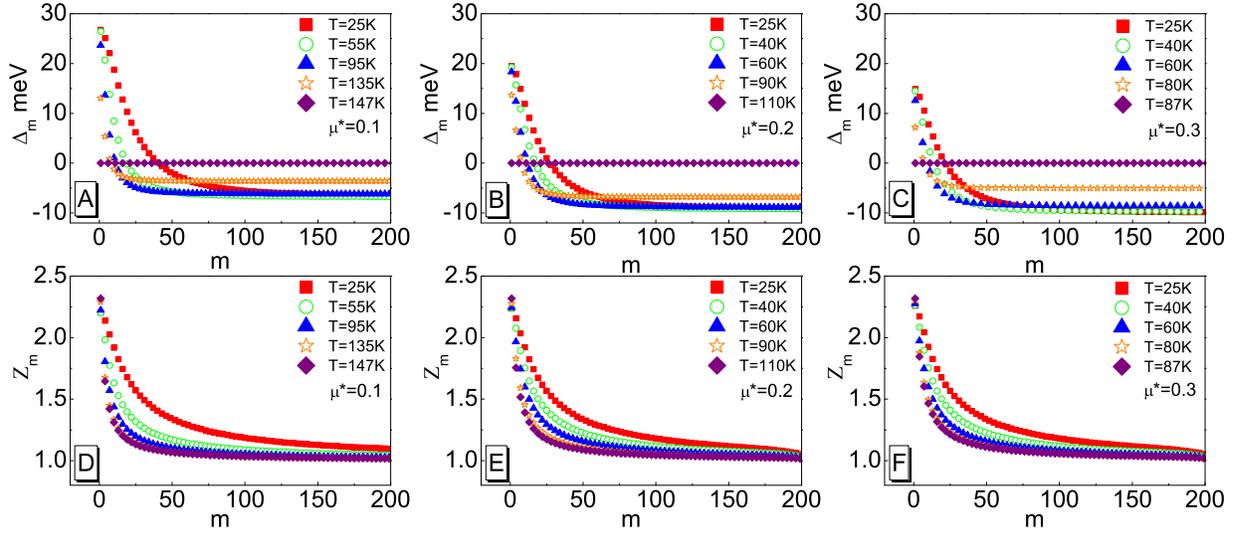}
\caption{(A)-(C) The order parameter and (D)-(F) the wave function renormalization factor for the selected values of the temperature and the Coulomb pseudopotential. The first $200$ values of $\Delta_{m}$ and $Z_{m}$ have been presented.}
\label{f1}
\end{figure*}
%

\vspace*{0.25cm}

In the first step, basing on the numerical analysis, the possible range of the critical temperature has been determined. It has been found that $T_{C}$ varies in the range from $147$ K to $87$ K for $\mu^{\star}\in\left<0.1,0.3\right>$. Thus, regardless of the physical value of the parameter $\mu^{\star}$, in the $\rm{B_{2}H_{6}}$ compound, the occurrence of the high temperature superconducting state is highly expected.

It is worth noting that the value of the critical temperature can be determined by the McMillan or Allen-Dynes formula \cite{McMillan}, \cite{AllenDynes}. However, the obtained results are significantly depressed in comparison to the results obtained directly from the Eliashberg equations. 

The forms of the order parameter and the wave function renormalization factor on the imaginary axis have been presented in \fig{f1}. Basing on the obtained results, it has been found that the order parameter values strongly decrease with the increasing temperature and the Coulomb pseudopotential. 
In turn, the wave function renormalization factor is much less dependent on $T$ and $\mu^{\star}$.  

%
\begin{figure}[ht]
\includegraphics[width=\columnwidth]{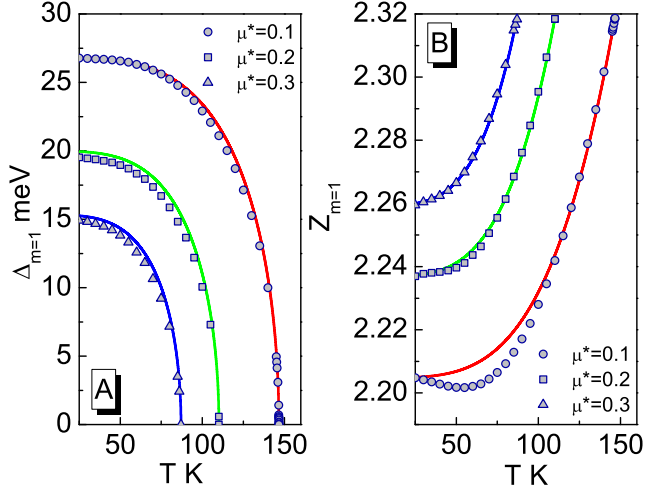}
\caption{(A) The order parameter $\Delta_{m=1}$ and (B) the renormalization factor $Z_{m=1}$ as a function of the temperature for selected values of the Coulomb pseudopotential. The symbols represent the exact numerical results. The solid lines have been plotted on the basis of the formulas (6) and (7).}
\label{f2}
\end{figure}
%

The complete dependence of the order parameter and the renormalization factor on the temperature and the Coulomb pseudopotential can be most conveniently traced after plotting the functions $\Delta_{m=1}\left(T,\mu^{\star}\right)$ and $Z_{m=1}\left(T,\mu^{\star}\right)$. The obtained results have been shown in \fig{f2}. 

It should be noted that from the physical point of view, the quantity  $2\Delta_{m=1}\left(T,\mu^{\star}\right)$ approximately determines the value of the energy gap at the Fermi level. On the other hand, the function $Z_{m=1}\left(T,\mu^{\star}\right)$ is related to the effective mass of the electron. 
In particular: $m^{\star}_{e}\simeq Z_{m=1}\left(T,\mu^{\star}\right)m_{e}$, where the symbol $m_{e}$ represents the electron band mass. 

Let us note that the numerical results presented in \fig{f2} can be parametrized by the means of the following formulas:
\begin{equation}
\label{r6}
\Delta_{m=1}\left(T,\mu^{\star}\right)=\Delta_{m=1}\left(\mu^{\star}\right)\sqrt{1-\left(\frac{T}{T_{C}}\right)^{\Gamma}}, 
\end{equation}
and
\begin{eqnarray}
\label{r7}
Z_{m=1}\left(T,\mu^{\star}\right)&=&Z_{m=1}\left(\mu^{\star}\right)\\ \nonumber 
&+&\left[Z_{m=1}\left(T_{C}\right)-Z_{m=1}\left(\mu^{\star}\right)\right]\left(\frac{T}{T_{C}}\right)^{\Gamma},
\end{eqnarray}
where $\Gamma=3.75$. The functions $\Delta_{m=1}\left(\mu^{\star}\right)$ and $Z_{m=1}\left(\mu^{\star}\right)$ can be written as: 
\begin{equation}
\label{r8}
\Delta_{m=1}\left(\mu^{\star}\right)=108.6\left(\mu^{\star}\right)^{2}-100.9\mu^{\star}+35.8,
\end{equation}
and
\begin{equation}
\label{r9}
Z_{m=1}\left(\mu^{\star}\right)=-0.484\left(\mu^{\star}\right)^{2}+0.467\mu^{\star}+2.163.
\end{equation}
The value $Z_{m=1}\left(T_{C}\right)$ need to be calculated on the basis of the expression: $Z_{m=1}\left(T_{C}\right)=1+\lambda=2.32$, where  
${\lambda\equiv 2\int^{\Omega_{\rm{max}}}_0 \alpha^2\left(\Omega\right)F\left(\Omega\right)/\Omega}$.

\vspace*{0.25cm}

On the basis of the solutions of the Eliashberg equations, the free energy difference between the superconducting and normal state has been calculated \cite{Bardeen}: 
\begin{eqnarray}
\label{r10}
\frac{\Delta F}{\rho\left(0\right)}&=&-\frac{2\pi}{\beta}\sum_{n=1}^{M}
\left(\sqrt{\omega^{2}_{n}+\Delta^{2}_{n}}- \left|\omega_{n}\right|\right)\\ \nonumber
&\times&(Z^{S}_{n}-Z^{N}_{n}\frac{\left|\omega_{n}\right|}
{\sqrt{\omega^{2}_{n}+\Delta^{2}_{n}}}).
\end{eqnarray}  
The symbol $\rho\left(0\right)$ denotes the value of the electron density of states at the Fermi level; $Z^{S}_{n}$ and $Z^{N}_{n}$ are the wave function renormalization factors for the superconducting ($S$) and the normal state ($N$), respectively. 

\begin{figure}[ht]
\includegraphics[width=\columnwidth]{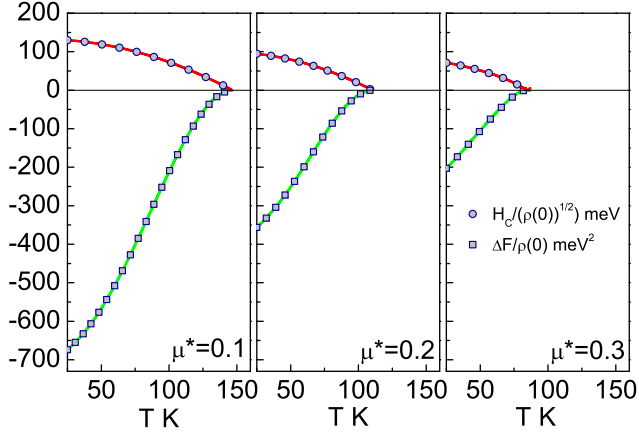}
\caption{(Lower panels) The dependence of the free energy on the temperature for the selected values of the Coulomb pseudopotential. 
(Upper panels) The thermodynamic critical field as a function of the temperature.}
\label{f3}
\end{figure}

The obtained results have been shown in the lower panels in \fig{f3}. It is easy to note that the increase of the Coulomb pseudopotential results in the strong decrease of the free energy value. In particular: 
$\left[\Delta F\left(T_{0}\right)\right]_{\mu^{\star}=0.3}/\left[\Delta F\left(T_{0}\right)\right]_{\mu^{\star}=0.1}=0.30$. 

%
%

In the next step, the thermodynamic critical field has been determined:
\begin{equation}
\label{r11}
\frac{H_{C}}{\sqrt{\rho\left(0\right)}}=\sqrt{-8\pi\left[\Delta F/\rho\left(0\right)\right]}.
\end{equation}

The results have been shown in the upper panels in \fig{f3}. The destructive impact of the Coulomb pseudopotential on the value of the thermodynamic critical field has been determined as: $\left[H_{C}\left(0\right)\right]_{\mu^{\star}=0.3}/\left[H_{C}\left(0\right)\right]_{\mu^{\star}=0.1}=0.55$, where 
$H_{C}\left(0\right)\equiv H_{C}\left(T_{0}\right)$.

The specific heat of the superconducting state ($C^{S}$) has been calculated on the basis of the formula:
\begin{equation}
\label{r12}
C^{S}=C^{N}+\Delta C,
\end{equation}
where the specific heat for the normal state ($C^{N}$) is the linear function of the temperature: 
$\frac{C^{N}\left(T\right)}{k_{B}\rho\left(0\right)}=\frac{\gamma}{\beta}$. The Sommerfeld constant is equal to: $\gamma\equiv\frac{2}{3}\pi^{2}\left(1+\lambda\right)$. The difference in the specific heat between the superconducting and normal state has been determined using the formula: 
\begin{equation}
\label{r13}
\frac{\Delta C\left(T\right)}{k_{B}\rho\left(0\right)}=-\frac{1}{\beta}\frac{d^{2}\left[\Delta F/\rho\left(0\right)\right]}{d\left(k_{B}T\right)^{2}}.
\end{equation}
%

%
\begin{figure}[ht]
\includegraphics[width=\columnwidth]{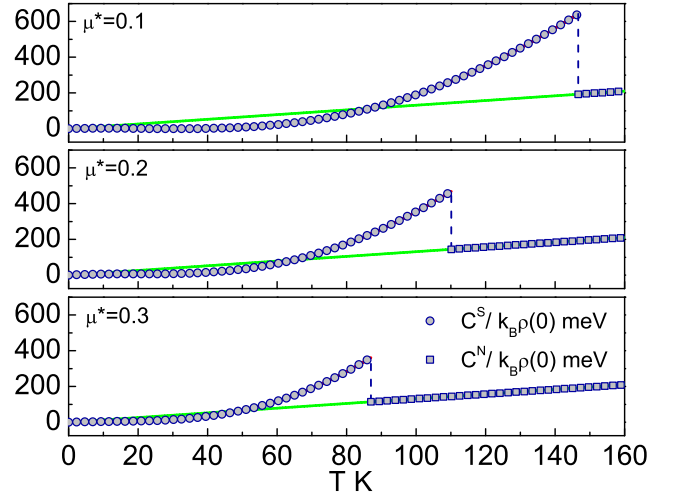}
\caption{
The specific heat of the superconducting state and the normal state as a function of the temperature for the selected values of the Coulomb pseudopotential.}
\label{f4}
\end{figure}
%

\fig{f4} presents the dependence of specific heats on the temperature and the Coulomb pseudopotential. It can be seen that at the critical temperature exists a characteristic jump, which is marked by the vertical dashed line. In addition, it should be also noted that the value of the specific heat jump  strongly decreases with the increasing of Coulomb pseudopotential. In particular, the ratio 
$\left[\Delta C\left(T_{C}\right)\right]_{\mu^{\star}=0.3}/\left[\Delta C\left(T_{C}\right)\right]_{\mu^{\star}=0.1}$ equals $0.55$. 

Basing on the obtained results, the values of the two characteristic dimensionless ratios have been calculated: 
\begin{equation}
\label{r14}
R_{H}\equiv \frac{T_{C}C^{N}\left(T_{C}\right)}{H^{2}_{C}\left(0\right)}, \qquad {\rm and} \qquad 
R_{C}\equiv \frac{\Delta C\left(T_{C}\right)}{C^{N}\left(T_{C}\right)}.
\end{equation}

It has been found that in the range of the considered values of the Coulomb pseudopotential, the parameter $R_{H}$ increases ($R_{H}\in\left<0.144,0.168\right>$) with the increasing value of $\mu^{\star}$, whereas the parameter $R_{C}$ decreases ($R_{C}\in\left<2.33,2.17\right>$). It should be boldly underlined that in the framework of the BCS theory, the ratios $R_{H}$ and $R_{C}$ are the universal constants: $\left[R_{H}\right]_{\rm BCS}=0.168$ and $\left[R_{C}\right]_{\rm BCS}=1.43$ \cite{BCS}. The difference between the Eliashberg predictions and the BCS theory results from the existence of the strong-coupling and retardation effects, which appear in the $\rm{B_{2}H_{6}}$ compound.
  
\begin{figure*}[ht]
\includegraphics[scale=0.60]{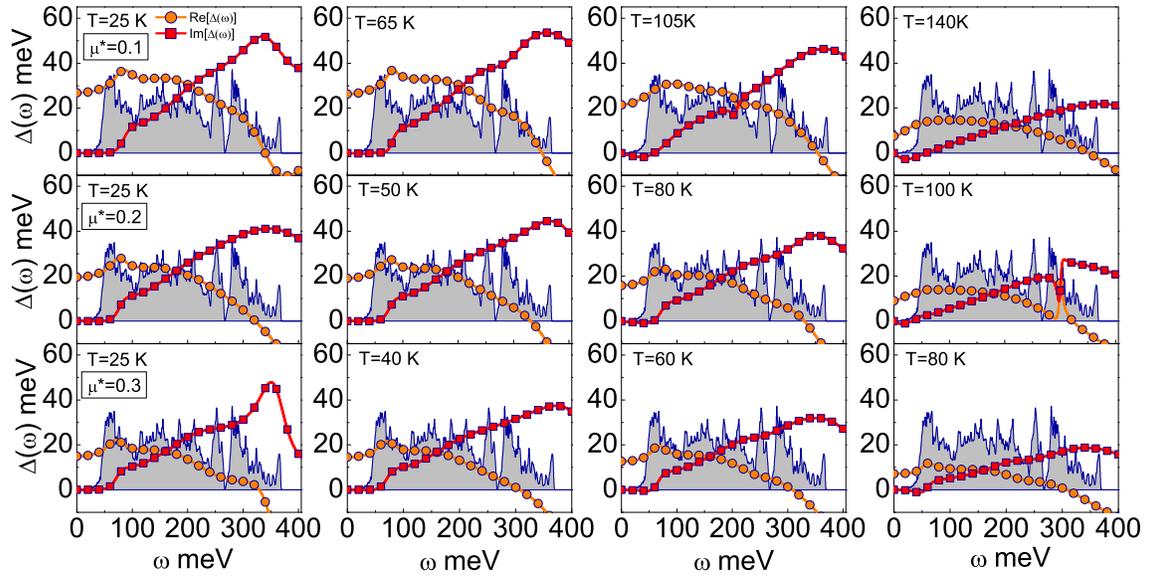}
\caption{ The order parameter on the real axis for the selected values of the temperature and the Coulomb pseudopotential. 
Additionally, the rescaled Eliashberg function has been plotted ($70\alpha^{2}F\left(\Omega\right)$).}
\label{f5}
\end{figure*}

By using the functions $\phi_{n}$ and $Z_{n}$, one can solve the Eliashberg equations in the mixed representation. The obtained results for the order parameter have been shown in \fig{f5}.

%
%
%
\begin{figure*}[ht]
\includegraphics[scale=0.60]{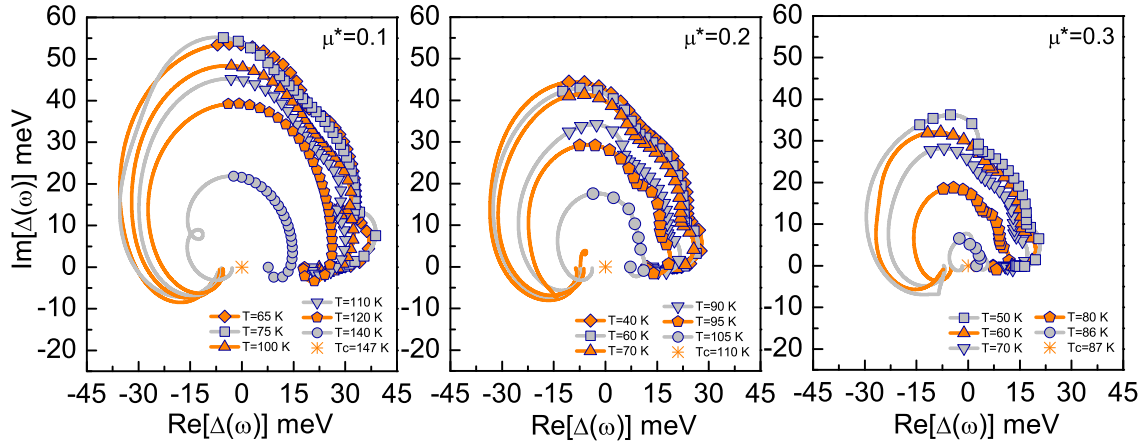}
\caption{ The order parameter on the complex plane for the selected values of the temperature and the Coulomb pseudopotential. The lines with the symbols have been obtained for $\omega\in\left<0,\Omega_{{\rm max}}\right>$; the lines without the symbols correspond to the range of the frequencies from 
$\Omega_{{\rm max}}$ to $\omega_{c}$.}
\label{f6}
\end{figure*}
%

It has been found that for the low frequencies the non-zero value takes only the real part of the function $\Delta\left(\omega\right)$. This result indicates no damping effects \cite{Varelogiannis}. Let us note that the imaginary part of the order parameter becomes non-zero only for the higher frequencies.

In addition, it can be easily seen that the functions ${\rm Re}\left[\Delta\left(\omega\right)\right]$ 
and ${\rm Im}\left[\Delta\left(\omega\right)\right]$ have a rather smooth course. However, the existing weak maximums or minimums are related to the corresponding peaks in the Eliashberg function. 

It also seems worthy to plot the values of the order parameter on the complex plane (see \fig{f6}). In the present case, the characteristic shapes of the deformed spirals can be observed. It should be noted that the radius of the considered curves becomes strongly reduced with the increasing temperature and the Coulomb pseudopotential. 

The curves presented in \fig{f6} allow in an easy way to characterize the effective potential for the electron-electron interaction. In the case when the curve meets the condition ${\rm Re\left[\omega\right]>0}$, the potential is pairing \cite{Varelogiannis}. Thus, on the basis of \fig{f6}, it has been found that the effective interaction between electrons leads to the formation of the superconducting condensate in the frequencies range from $0$ to 
$\sim 0.90\Omega_{\rm max}$, if $\mu^{\star}=0.1$. The increase in the value of the Coulomb pseudopotential causes the narrowing of the considered range of $\omega$ from the side of the higher frequencies. 

The energy gap at the Fermi level has been determined on the basis of the equation: 
\begin{equation}
\label{r15}
\Delta\left(T\right)={\rm Re}\left[\Delta\left(\omega=\Delta\left(T\right)\right)\right].
\end{equation}

From the physical point of view, the most interesting is the value for the lowest temperature: $2\Delta\left(0\right)$, where $\Delta\left(0\right)\equiv\Delta\left(T_{0}\right)$. As a result of the calculations the following has been obtained:  
$2\Delta\left(0\right)\in\left<26.77,14.91\right>$ meV for $\mu^{\star}\in\left<0.1,0.3\right>$. 

%
\begin{figure}[ht]
\includegraphics[width=\columnwidth]{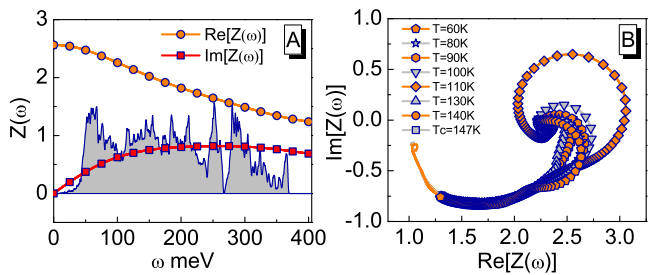}
\caption{(A) The wave function renormalization factor on the real axis. Additionally, the rescaled Eliashberg function has been plotted ($3\alpha^{2}F\left(\Omega\right)$). (B) The wave function renormalization factor on the complex plane. The lines with the symbols have been obtained for $\omega\in\left<0,\Omega_{{\rm max}}\right>$; the lines without the symbols correspond to the range of the frequencies from $\Omega_{{\rm max}}$ to $\omega_{c}$.}
\label{f7}
\end{figure}
%

On the basis of the results above, it is very easy to estimate the value of the dimensionless parameter:
\begin{equation}
\label{r16}
R_{\Delta}\equiv \frac{2\Delta\left(0\right)}{k_{B}T_{C}}.
\end{equation}

The obtained result has the form: $R_{\Delta}\in\left<4.24,3.98\right>$. It should be noted that in the framework of the BCS theory the ratio $R_{\Delta}$ is the universal constant and it is equal to $3.53$ \cite{BCS}. Hence, it can be easily noticed that the BCS model is too simple to predict in the correct way the physical value of $R_{\Delta}$ for the $\rm{B_{2}H_{6}}$ compound under the pressure at $360$ GPa.

The wave function renormalization factor on the real axis for $T=T_{C}$ and $\mu^{\star}=0.1$ has been shown in \fig{f7} (A). Next, we have presented the courses of the renormalization factor on the complex plane for the selected values of the temperature. 

The second solution of the Eliashberg equations allowed us to calculate the exact value of the electron effective mass: 
$m^{\star}_{e}=Z\left(\omega=0\right)m_{e}$. As a result, it has been found that in the entire temperature range, from $T_{0}$ to $T_{C}$ the electron effective mass is large and it reaches its maximum value at the critical temperature ($\left[m_{e}\right]_{\rm max}=2.56$).


\vspace*{0.25cm}

In summary, in the paper we have determined all relevant thermodynamic parameters of the superconducting state in the $\rm{B_{2}H_{6}}$ compound. The pressure of $360$ GPa has been taken into account. 

It has been found that, regardless assumed value of the Coulomb pseudopotential, the critical temperature is high. In particular: $T_{C}\in\left<147,87\right>$ K, for $\mu^{\star}$ changing in the range from $0.1$ to $0.3$. 

Other thermodynamic parameters differ significantly from the predictions of the classical BCS theory. This is especially visible for the low values of the Coulomb pseudopotential. In particular, it has been proven that the dimensionless ratios, characterizing the derogation from the results of the BCS theory, take the values: $R_{\Delta}\in\left<4.24,3.98\right>$, $R_{C}\in\left<2.33,2.17\right>$, and $R_{H}\in\left<0.144,0.168\right>$.

\begin{acknowledgments}

The authors would like to thank Prof. K. Dzili{\' n}ski for providing excellent working conditions and the financial support.

Additionally, we are grateful to the Cz{\c{e}}stochowa University of Technology - MSK
CzestMAN for granting access to the computing infrastructure built in the
project No. POIG.02.03.00-00-028/08 "PLATON - Science Services Platform".

\end{acknowledgments}
%
%

%
\end{document}